\documentclass[reprint,aps,prl,amsmath,amssymb]{revtex4-2}

\usepackage[version=4]{mhchem}

\usepackage{siunitx}
\usepackage{graphicx}% Include figure files
\usepackage{balance}

\usepackage{dcolumn}% Align table columns on decimal point
\usepackage{bm}% bold math
\begin{document}

\preprint{}

\title{\textbf{Heavy Fermion Phase Diagram in Magic-angle Twisted Trilayer Graphene} 
}% 

\author{%
Le Zhang\textsuperscript{1,2,*},
Wenqiang Zhou\textsuperscript{1},
Xinjie Fang\textsuperscript{1},
Zhen Zhan\textsuperscript{5},
Kenji Watanabe\textsuperscript{6},
Takashi Taniguchi\textsuperscript{7},
Yi-feng Yang\textsuperscript{3,4},
and Shuigang Xu\textsuperscript{1,2,}
}
\thanks{ zhangle@westlake.edu.cn}
\thanks{ xushuigang@westlake.edu.cn}

\affiliation{%
\textsuperscript{1}Key Laboratory for Quantum Materials of Zhejiang Province, Department of Physics, School of Science, Westlake University, Hangzhou, China\\
\textsuperscript{2}Institute of Natural Sciences, Westlake Institute for Advanced Study, Hangzhou, China\\
\textsuperscript{3}Beijing National Laboratory for Condensed Matter Physics, Institute of Physics, Chinese Academy of Sciences, Beijing, China\\
\textsuperscript{4}University of Chinese Academy of Sciences, Beijing, China\\
\textsuperscript{5}IMDEA Nanociencia, Madrid, Spain\\
\textsuperscript{6}Research Center for Electronic and Optical Materials, National Institute for Materials Science, Tsukuba, Japan\\
\textsuperscript{7}Research Center for Materials Nanoarchitectonics, National Institute for Materials Science, Tsukuba, Japan
}

\date{\today}% It is always \today, today,

\begin{abstract}
The interplay between localized magnetic moments and itinerant electrons gives rise to exotic quantum states in condensed matter systems. Here, we demonstrate an electrically tunable heavy fermion phase diagram in magic-angle twisted trilayer graphene, achieved by controlling the Kondo hybridization between localized flat-band electrons and itinerant Dirac electrons via a displacement field. Our results reveal a continuous quantum phase transition from an antiferromagnetic semimetal to a paramagnetic heavy fermion metal. At quantum critical point, we observe effective mass divergence and Fermi surface reconstruction. This highly tunable platform offers unprecedented control over heavy fermion physics, establishing moiré heterostructures as a versatile arena for exploring correlated quantum phases—including potential unconventional superconductivity—in two-dimensional limit.

\end{abstract}

\maketitle

%\tableofcontents

~ Two-dimensional (2D) moiré superlattices with flat bands have emerged as a versatile platform for exploring strongly correlated matters \cite{Checkelsky2024}. The strong on-site interactions among localized electrons can give rise to a plethora of quantum phases, including unconventional superconductivity \cite{Cao2018SC,Lu2019}, correlated insulators \cite{Cao2018CI}. Beyond the established Hubbard physics, moiré materials can be engineered to simulate Kondo lattice physics by coupling localized moments ($f$-electrons) with a sea of itinerant conduction electrons ($c$-electrons) \cite{Chou2023,Kumar2022,Guerci2023}. In magic-angle twisted bilayer graphene (MATBG), hybridization between $f$-electrons centered at AA-stacking regions and plane-wave-like itinerant $c$-electrons can produce topological flat bands near the Fermi level $E_\text{f}$ \cite{Ghosh2025,Merino2025,Pierce2025,Song2022,Zhou2024,Hu2023,Lau2025}. Its counterpart, magic-angle twisted trilayer graphene (MATTG), hosts both flat bands and a set of Dirac bands. These additional Dirac bands not only help stabilize superconducting states \cite{Park2021SC,Lin2022,Hao2021}, but also provide extra $c$-electrons and an unprecedented form of $in\mbox{-}situ$ tunability. This enables the emulation of the full heavy-fermion phase diagram, going beyond what is possible in MATBG and semiconducting transition-metal dichalcogenide heterobilayers \cite{Zhao2023,Zhao2024,Ramires2021,Yu2023}.

~ In conventional rare-earth-based Kondo lattices, the ground state can evolve into a heavy Fermi liquid via the Kondo effect or develop magnetic order mediated the Ruderman-Kittel-Kasuya-Yosida (RKKY) interaction. In MATTG, the competing interplay can be continuously tuned via a perpendicular displacement field. Here, we report the experimental observation of a continuous crossover from a heavy Fermi liquid to an antiferromagnetic semimetal, along with direct evidence of Fermi surface reconstruction at filling factor $\nu=n/n_0=3$ (here $n_0$ is the charge density corresponding to one electron per moiré unit cell). By mapping the generalized Doniach phase diagram in a single dual-gated device, our work may illuminate the mechanism behind unconventional superconductivity and reveal novel emergent phenomena beyond those found in bulk Kondo lattice materials.

~ The high-quality dual-graphite-gated MATTG devices, featuring three adjacent graphene layers twisted in an alternating sequence as illustrated in Fig. 1(a), are prepared by the dry transfer method. The vertical mirror symmetry structure gives rise to a unique band structure at zero displacement field $D=0$ \si{V~nm^{-1}}, comprising MATBG-like flat bands and Dirac bands originated from the additional monolayer graphene [Fig. 1(b)]. Upon applying a large $D=0.9$ \si{V~nm^{-1}}, the flat bands hybridize with the Dirac bands, leading to the coupling of $f$-electrons with $c$-electrons and the formation of a Kondo lattice [Fig. 1(c), right]. To experimentally probe this Kondo phase, we have measured two MATTG devices with twist angles of 1.52° (device D1) and 1.45° (device D2). Despite the slight variation in twist angle, the electronic properties of both devices show high consistency (see Supplemental Material Fig. S1 \cite{SM}). Following established transport characterization \cite{Park2021SC,Hao2021}, we first identify the electronic phases in MATTG. Figure 1(d) displays a typical longitudinal resistance $R_{xx}$ mapping as a function of filling factor $\nu$ and displacement field $D$ at $T=1.8$ K, where $\nu$ and $D$ are controlled by top/bottom gate voltages. The correlated insulating resistance peaks at integer filling $\nu = 1, \pm 2, 3$ and charge density wave peaks at fractional fillings become more pronounced as the magnitude of $|D|$ increases and eventually evolve into low-resistance metallic states beyond a critical displacement field.

\begin{figure}[!htbp]
\includegraphics[width=\columnwidth]{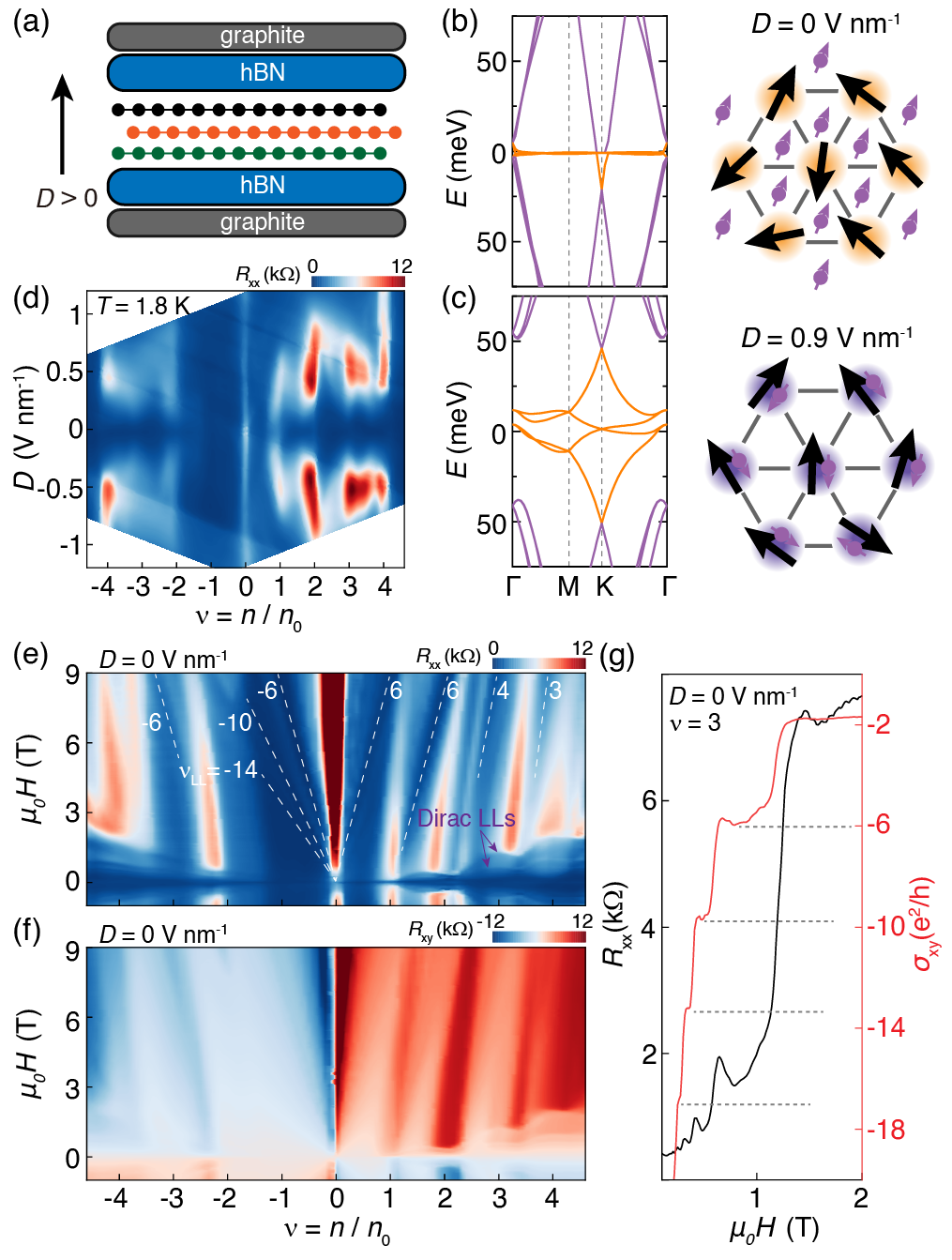}% Here is how to import EPS art
\caption{\label{fig:fig1} (a) Schematic of the dual-gated MATTG device. (b), (c) Tight-binding band structures for MATTG at zero (b) and finite displacement field (c). Right panel, schematic of localized moments (black arrow) interacting with itinerant electrons (purple arrow) at $D=0$ \si{V~nm^{-1}} and $D=0.9$ \si{V~nm^{-1}}. (d) Longitudinal resistance $R_{xx}$ as a function of $\nu$ and $D$ at $T=1.8$ K. (e), (f) Landau fan diagrams of $R_{xx}$ (e) and $R_{xy}$ (f) at $D=0$ \si{V~nm^{-1}}. (g) Quantum Hall states at $D=0$ \si{V~nm^{-1}} and $\nu=3$. Gray dashed lines mark quantized values corresponding to the sequence -2, -6, -10, \ldots}
\end{figure}

~ The coexistence of $f$- and $c$-electrons is further evident in the Landau fan diagram. As shown in Fig.1(e) and (f), the decoupled flat bands and Dirac bands at zero $D$ give rise to two distinct sets of Landau fans: dense MATBG-like fans originating from integer and fractional fillings at high magnetic field and extra quantum oscillations emanating from the charge neutrality point (CNP) at low field. By measuring the quantum Hall states [Fig. 1(g)], we can confirm that the additional quantum oscillations are in fact from Dirac bands. From the temperature dependence of Shubnikov-de Haas (SdH) oscillations at low magnetic field, we extract an effective mass of $m^* \sim 0.005m_0$ ($m_0$ is the free electron mass), consistent with massless Dirac fermions in monolayer graphene (Supplemental Material Fig. S7 \cite{SM}) \cite{Novoselov2005,Zhang2005}.

\begin{figure*}[!t]
\includegraphics[width=\textwidth]{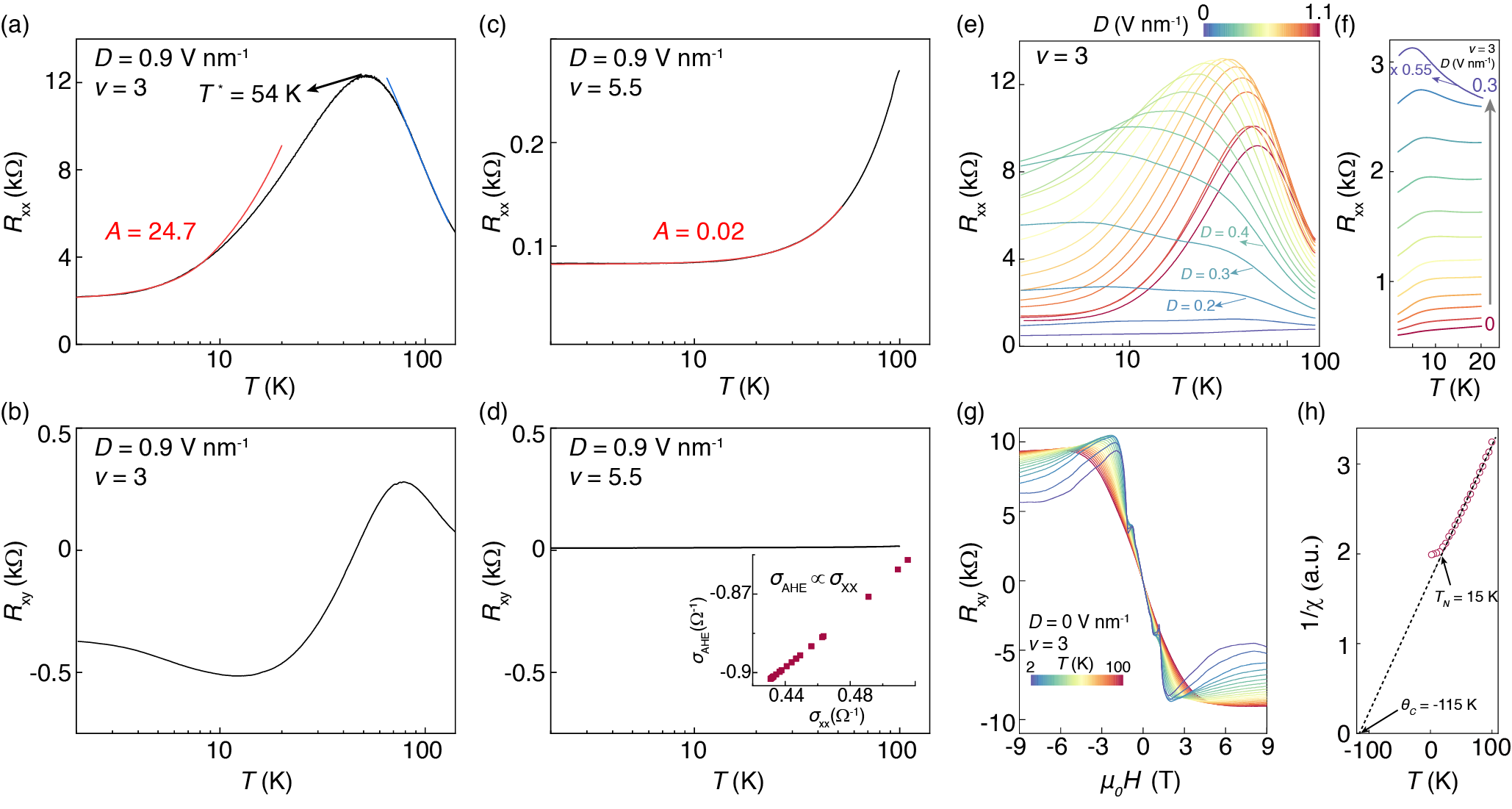}% Here is how to import EPS art
\caption{\label{fig:fig2}(a) $R_{xx}$ versus temperature at $D=0.9$ \si{V~nm^{-1}} and $\nu=3$. The blue line represents the $\ln T$ fit at high temperature regime, while the red line corresponds the $T^2$ fit at low temperature regime. (b) $R_{xy}\mbox{-}T$ curves at $\nu=3$ for $D=0.9$ \si{V~nm^{-1}}. (c),(d) $R_{xx}\mbox{-}T$ (c) and $R_{xy}\mbox{-}T$ (d) curves obtained beyond the flat bands at $D=0.9$ \si{V~nm^{-1}} and $\nu=5.5$. Inset: The linear dependence of $\sigma_{\text{AHE}}$ versus $\sigma_{xx}$ for $D=0$ \si{V~nm^{-1}} and $\nu=3$. (e) $R_{xx}\mbox{-}T$ curves at $\nu=3$ for various displacement fields. (f) $R_{xx}\mbox{-}T$ curves at $\nu=3$ below 20 K with small displacement field ranging from $D=0$ \si{V~nm^{-1}} to $D=0.3$ \si{V~nm^{-1}}. (g) Magnetic-field-dependent $R_{xy}$ for $D=0$ \si{V~nm^{-1}} and $\nu=3$ at various temperatures. (h) Temperature dependence of the inverse magnetic susceptibility for $D = 0\text{ V nm}^{-1}$ and $\nu = 3$.}
\end{figure*}

~ Overall, the hallmarks of gate-tunable correlated insulators at low temperature are highly reproducible (see Supplemental Material Fig. S1 \cite{SM}) and consistent with prior experimental results \cite{Park2021SC,Hao2021,Shen2023,Liu2022}. Intriguingly, we also observe counterintuitive resistance behaviors at high temperatures. Figure 2(a) presents the temperature-dependent longitudinal resistance $(R_{xx}\mbox{-}T)$ curve at $D=0.9$ \si{V~nm^{-1}} and $\nu=3$. In the high temperature regime $T>54$ \si{K}, the $R_{xx}\mbox{-}T$ curve exhibits a logarithmic temperature dependence, a hallmark signature of spin-flip scattering via Kondo effect \cite{Chen2011}. At an intermedia temperature $T\approx54$ K, the resistance reaches a maximum, signaling the onset of coherence, where localized moments start to be screened by itinerant $c$-electrons. Below this temperature, the resistance displays typical metallic behavior and transition into a Fermi liquid regime described by $R_{xx}=R_0+AT^2$ at $T<10$ \si{K}, where $R_0$ is the residual resistance, and $A^{0.5}$ is linearly proportional to the quasiparticle effective mass $m^*$ according to Kadowaki-Woods scaling \cite{Jacko2009,Kadowaki1986}. For comparison, we also measured the $R_{xx}\mbox{-}T$ curve obtained at $D=0.9$ \si{V nm^{-1}} and $\nu=5.5$ [Fig. 2(c)], where only the Dirac bands and higher-energy dispersive bands are involved. As expected, the $R_{xx}\mbox{-}T $ curve exhibits pure Fermi liquid behavior with an extracted coefficient $A$ to be 0.02 $\Omega/\mathrm{K}^2$. The corresponding $m^*$ was determined by the temperature dependence of SdH oscillations, yielding a small value of $0.06 m_0$ (see Supplemental Material Fig. S8 \cite{SM}). Accordingly, we can estimate that the effective mass for $\nu=3$, $D=0.9$ \si{V~nm^{-1}} is enhanced by 33 times, yielding $m^*\sim2m_0$—consistent with previous results in graphene moiré superlattices \cite{Cao2018CI,Park2021SC}. The standard $R_{xx}-T$ curve and dramatic enhancement of effective mass strongly indicate the emergence of a heavy fermion state for $\nu=3$ at high displacement field. We also compare the temperature dependent Hall resistance $R_{xy}-T$ at $\nu=3$ and $\nu=5.5$ for $D=0.9$ \si{V~nm^{-1}}. While  $R_{xy}\mbox{-}T$ curve for $\nu=5.5$ is nearly temperature-independent [as shown in Fig. 2(d)], the corresponding curve for $\nu=3$ exhibits characteristic heavy fermion behavior:  $R_{xy}$ reaches a maximum at 77 \si{K}, changes sign at 45 \si{K}, passes through a minimum at 13 \si{K}, and then increases gradually with the expansion of Fermi surface at low temperature \cite{Paschen2005,31}.

~ Having established the presence of heavy fermion, we now investigate the evolution of Kondo hybridization with displacement field. Figure 2(e) and 2(f) display $R_{xx}\mbox{-}T$ curves at $\nu=3$ with displacement fields ranging from 0 \si{V~nm^{-1}} to 1.1 \si{V~nm^{-1}}. As shown in Fig. 2(f), the $R_{xx}\mbox{-}T$ curve displays semimetallic behavior with an abrupt resistance drop at low temperature, which we attribute to the suppression of spin-disorder scattering \cite{32}. By analyzing the magnetic field dependence of both $R_{xx}$ and  $R_{xy}$ [Fig. 3(c) and Fig. 2(e), respectively], we observe clear signatures of spin-flop transition magnetoresistance (MR) \cite{32,33,34,35} and a giant anomalous Hall effects (AHE) \cite{35}. 
From the AHE measurements in Fig. 2(g), we can extract the magnetic susceptibility as:
\begin{equation*}
\chi \approx \left. \frac{\partial R_{xy}}{\partial H} \right|_{H \to 0}.
\end{equation*}
As shown in Fig.~2(h), the temperature dependence of $1/\chi$ follows the Curie-Weiss law:
\begin{equation*}
\chi = \frac{C}{T - \theta_C}.
\end{equation*}
The linear fit yields a negative Curie-Weiss temperature, $\theta_C = -115\text{ K}$, indicating predominant antiferromagnetic interactions \cite{36}. The N{\'e}el temperature shown in Fig.~2(h) ($T_\text{N} \approx 15\text{ K}$), identified from the deviation from linearity in $1/\chi$, aligns well with the $T_\text{N} = 14.2\text{ K}$ determined from the minimum in the derivative $\mathrm{d}R_{xx}/\mathrm{d}T - T$ at same $D = 0\text{ V nm}^{-1}$ (Fig.~S12). The yielded $|\theta_C|$ much larger than $T_\text{N}$ indicates the competition between the RKKY interaction and the Kondo effect, which frequently occurs in heavy fermion systems \cite{37,38}. These features provide strong evidence for the emergence of an antiferromagnetic order at low displacement fields. The persistence of AHE up to 100 K is unlikely to originate from an intrinsic large Berry curvature \cite{39,40}. The linear relationship between $\sigma_\text{AHE}$ versus $\sigma_{xx}$, as shown in the inset of Fig. 2(d), suggests an extrinsic skew-scattering origin, where the $c$-electrons are asymmetrically scattered by localized moments \cite{41}.

~ By normalizing the $R_{xx}\mbox{-}T$ curve, we present the first 2D heavy fermion phase diagram in MATTG as shown in Fig. 3(a), featuring an antiferromagnetic semimetal and a paramagnetic heavy fermion state. Starting at $D=0$ \si{V~nm^{-1}}, the antiferromagnetic order is progressively suppressed by the displacement field, with the Néel temperature $T_\text{N}$ shifts toward zero. Beyond $D=0.4$ \si{V~nm^{-1}}, a new characteristic temperature $T^*$, signaling the onset of Kondo screening, rises proportionally with the displacement field, as does the heavy Fermi liquid temperature $T_{\text{HFL}}$. This tendency is consistently reproduced in device D2 (see Supplemental Material Fig. S11 \cite{SM}). Intriguingly, as shown in Fig. 3(b), the fitted coefficients $A$ for discrete displacement fields $D$ exhibit a divergent behavior upon approaching the critical displacement field $D\approx0.5$ \si{V~nm^{-1}}.

~ The continuous evolution from a semimetal to a heavy fermion metal, accompanied by the divergence of the $A$ coefficients and effective mass $m^*$ (see Supplemental Material Fig. S7 \cite{SM}), suggests the presence of a quantum critical point (QCP) at the zero-temperature limit \cite{42}. In this scenario, the Fermi surface transitions from a small to a large configuration, as illustrated in the inset of Fig. 3(b). In the low-displacement-field AFM semimetal phase, static Kondo screening is absent. This means the ground state is not a Kondo singlet, and fully developed Kondo resonances do not form. Consequently, only itinerant Dirac electrons contribute to the Fermi volume, resulting in a small Fermi surface. In contrast, in the high-displacement-field heavy fermion liquid phase, the Kondo singlet ground state gives rise to pronounced Kondo resonances, which must be incorporated into the Fermi volume. This leads to a correspondingly large Fermi surface.

~ To unveil this crossover, we measured the magnetic-field dependence of $R_{xx}$ and  $R_{xy}$ across various value of $D$ at $T=2$ \si{K} [Fig. 3(c) and Fig. 3(d)]. In the weak-coupling regime ($D<0.3$ \si{V~nm^{-1}}), the MR exhibits prominent SdH oscillations at low magnetic fields ($\mu_0 H<1$ \si{T}). These oscillations are suppressed as the displacement field approaches $D\approx0.3$ \si{V~nm^{-1}} , consistent with the observed Landau fan behavior (Supplemental Material Fig. S3 \cite{SM}). Owing to the incoherent Kondo scattering, localized moments do not contribute to the Fermi-surface formation, ultimately leading to magnetic order below $T_\text{N}$. As $D$ increases, the localized moments hybridize with $c$-electrons through the Kondo effect, forming composite fermions with a heavily renormalized effective mass. The resulting mass enhancement suppresses SdH oscillations. The composite fermions incorporate into the Fermi volume, driving Fermi surface expansion. The sign reversal of the low-field Hall coefficient $R_\text{H}$ further corroborates this transition, signaling a change in the dominant charge carriers from $c$-electrons to positively charged Kondo singlets.

\begin{figure}[!htbp]
\includegraphics[width=\columnwidth]{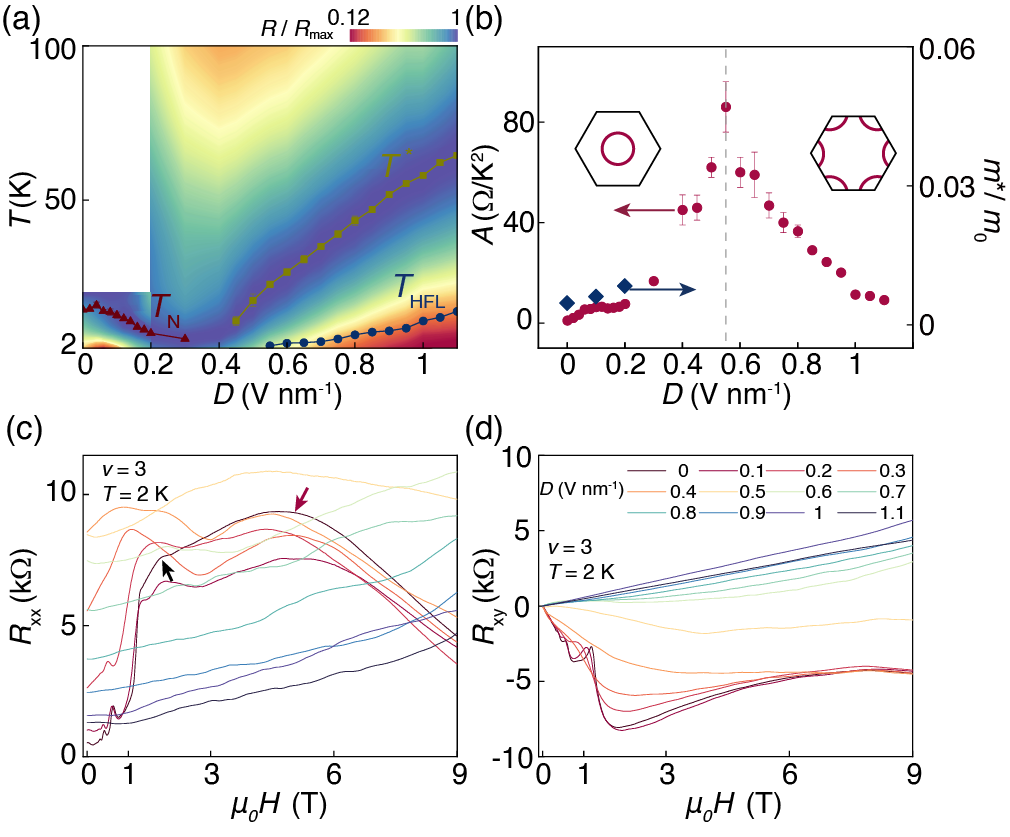}% Here is how to import EPS art
\caption{\label{fig:fig3} (a) $D$-tunable heavy fermion phase diagram. The Kondo screening temperatures $T^*$ and Néel temperature $T_\text{N}$ are determined from $\mathrm{d}R_{xx}/\mathrm{d}T$. The onset of heavy Fermi liquid, $T_\text{HFL}$, is identified when the $R_{xx}\mbox{-}T$ curve deviates from the Fermi-liquid $T^2$ dependence \cite{SM}. (b) The fitted $A$ coefficient of the $T^2$ term (red circles) and effective mass $m^*$ (blue diamonds) extracted from SdH oscillations at low-temperature regime as a function of $D$. Inset: Schematic illustration of the large Fermi surface (right) corresponding to the Kondo singlet ground state and the small Fermi surface (left) associated with the incompletely screened semimetal. (c), (d) Magnetic-field-dependent $R_{xx}$ (c) and $R_{xy}$ (d) for discrete $D$. The black and red arrows in (c) indicate the occurrence of spin-flop at 1.8 T and 5.1 T.}
\end{figure}

~ Although the band structure and Fermi surface evolution in a single-gate device can be directly imaged with a quantum twisting microscope \cite{43}, our dual-gated configuration necessitated a different approach. We therefore employed well-established indirect transport measurements to probe the Fermi surface changes comprehensively. Figure 4(f) shows the fast Fourier transform of $R_{xx}$($1/\mu_0 H$) of two representative displacement field, $D=0$ \si{V~nm^{-1}} and $D=0.77$ \si{V~nm^{-1}}, respectively. The oscillation frequencies reveal that the Fermi surface expands significantly as $D$ increases \cite{44}. Additionally, the effective mass undergoes a tenfold enhancement (Supplemental Material Fig. S10 \cite{SM}).

\begin{figure}[!htbp]
\includegraphics[width=\columnwidth]{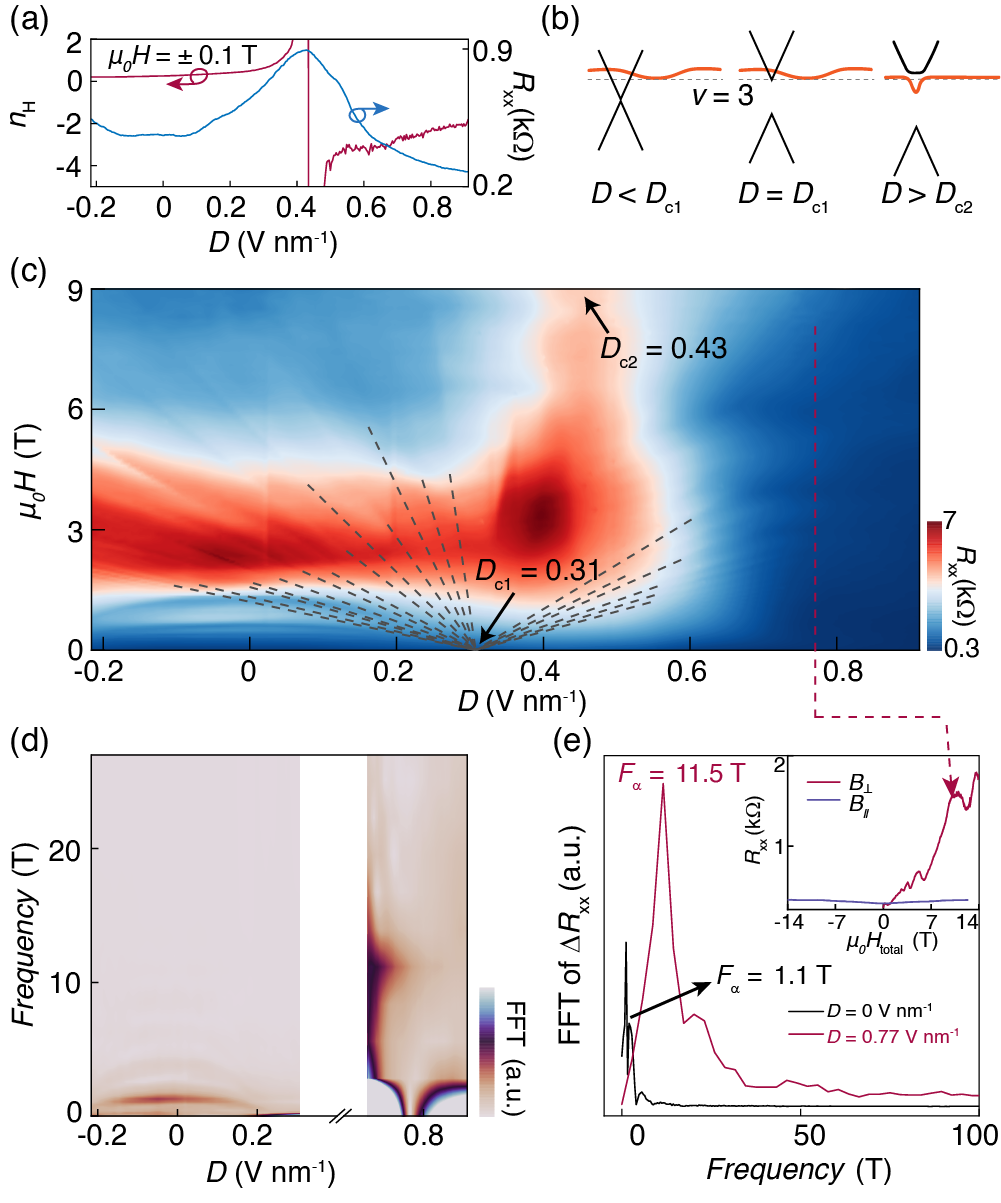}% Here is how to import EPS art
\caption{\label{fig:fig4} (a) Hall carrier density $n_H$ and longitudinal resistance $R_{xx}$ as a function of $D$ at $\nu=3$. (b) Schematic band structures near the critical displacement field. (c) $R_{xx}$ as a function of $D$ and $\mu_0 H$, taken at $\nu=3$ and $T=0.25$ K. The gray lines denote MIOs emanating from critical displacement field $D_{c1}=0.31$ \si{V~nm^{-1}}, red dashed line marks a representative heavy fermion oscillation at $D=0.77$ \si{V~nm^{-1}}. (d) Fast Fourier transform (FFT) of Landau fan, calculated from the data in (c). (e) FFT spectra of the SdH oscillation for $D=0$ \si{V~nm^{-1}} and $D=0.77$ \si{V~nm^{-1}}. Inset: MR under in-plane and out-of-plane magnetic field at $\nu=3$ and $D=0.77$ \si{V~nm^{-1}} at 0.25 K.}
\end{figure}

Furthermore, we measure the SdH oscillations by continuously scanning displacement at fixed magnetic fields \cite{45,46,47,48}. Figure 4(c) displays $R_{xx}$ at $\nu=3$ as a function of $\mu_0 H$ and $D$, measured in device D2 at $T=0.25$ K. In the ($D\mbox{-}\mu_0 H\mbox{-}R_{xx}$) phase space at low displacement field ($|D|<0.2 $ \si{V~nm^{-1}}) and magnetic field ($\mu_0 H<2$ T ), periodic low-resistance states (blue regions separated by white regions) emerge and are gradually suppressed as $D$ increased. Analysis of the SdH oscillations confirms that this periodicity arises from Dirac band $c$-electrons (Supplemental Material Fig. S10 \cite{SM}). At higher magnetic field ($\mu_0 H>2$ T ), a dense set of resistance oscillations emerge, which are originated from the scattering of charge carrier between minivalleys of flat and Dirac bands \cite{46,49}. Their trajectories, along with oscillations originating from higher $D$ side, converge at $D_{c1}=0.31$ \si{V~nm^{-1}} in the zero magnetic field limit [see the corresponding schematic in Fig.4(d)], signifying energy equilibrium between minivalleys at this critical displacement field. Similar magneto interminivalley oscillations (MIOs) can be observed at $\nu=2$ (Supplemental Material Fig. S13 \cite{SM}). At the vicinity of $D_{c2}=0.43$ \si{V~nm^{-1}}, we also observe a pronounced resistance peak and a divergence in Hall carrier density $n_H$, accompanied by a sign reversal [Fig. 4(a)], which are associated with Lifshitz transition in the Fermi surface topology.

~ Crucially, by tracking the SdH frequency as a function of $D$, we observe a clear Fermi surface reconstruction. In the weak-coupling regime, the SdH frequency shifts from 1.1 T to zero gradually. Near $D_{c2}=0.43$ \si{V~nm^{-1}}, however, the fermiology remains unclear due to the absence of well-defined SdH oscillations. The emergence of non-Fermi liquid behavior and nonlinear $I\mbox{-}V$ transport characteristics (Supplemental Material Fig. S11 \cite{SM}) suggests the possible absence of a well-defined Fermi surface in this regime, likely due to strong quantum fluctuations. Beyond the $D_{c2}=0.43$ \si{V~nm^{-1}}, a new frequency emerges at 11.5 T, reflecting a dramatic reconstruction of the Fermi surface driven by the strong Kondo effect between $f$- and $c$-electrons. The observed Fermi surface expansion, mass renormalization and divergence of the $A$ coefficients provide strong evidence of a QCP in the MATTG heavy fermion system \cite{42,44}.

\begin{figure}[!htbp]
\includegraphics[width=\columnwidth]{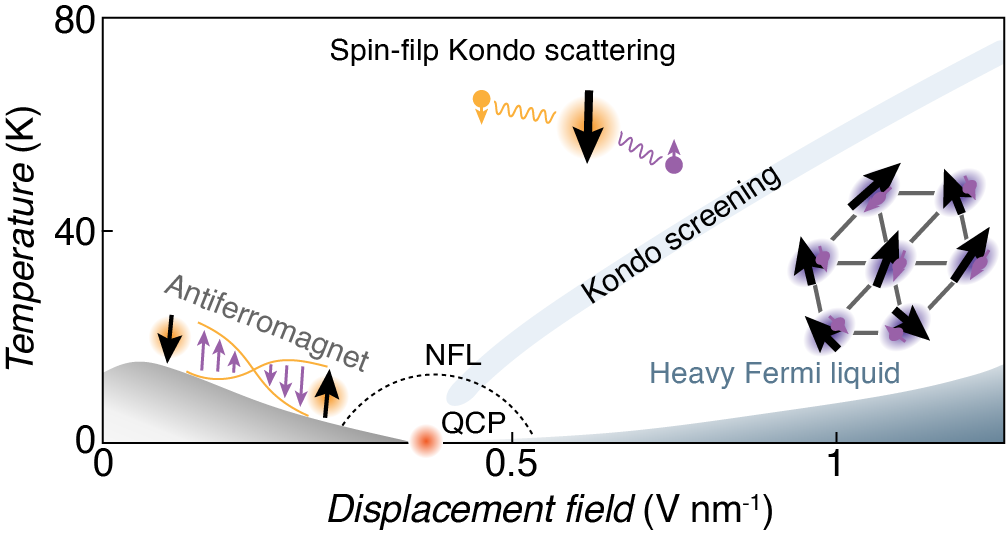}% Here is how to import EPS art
\caption{\label{fig:fig5} Phase diagram for heavy fermions in MATTG at $\nu=3$. The phase boundaries are schematized according to three characteristic temperatures: $T^*$, $T_\text{N}$, and $T_\text{HFL}$.}
\end{figure}

~ To better understand the fermiology evolution across different displacement field regimes, a schematic illustration is provided in Fig. 4(b). In MATTG, the displacement field modifies the band structure in two primary ways: ($\mathrm{i}$) hybridizing the flat and Dirac bands at $K/K'$ points, and ($\mathrm{ii})$ shifting the Dirac bands energetically \cite{50,51}. At $\nu=3$, the system remains semimetallic in the weak coupling regime ($D<D_{c1}$) due to the presence of Dirac bands. The Dirac bands shifted upward with displacement field increasing. At critical field $D_{c1}=0.31$ \si{V~nm^{-1}}, the Fermi level $E_\text{f}$ aligns with the Dirac point, as evidenced by the convergence of MIOs in the zero-field limit [Fig. 4(d)]. Upon further increasing displacement field beyond $D_{c2}=0.43$ \si{V~nm^{-1}}, the flat bands and Dirac bands undergo strong hybridization, leading to formation of new quasiparticle bands with ultra-flattened dispersions near $E_f$. This reconstruction of the electronic structure results in significant Fermi surface expansion, marking the onset of heavy Fermi liquid behavior at high displacement fields.

~ To summarize the evolution of versatile quantum states with displacement fields, Fig. 5 maps out a fully gate-tunable heavy fermion phase diagram in MATTG extracted from our experimental results. Crucially, $in\mbox{-}situ$ electric field tuning provides unprecedented resolution to access the QCP, revealing abrupt Fermi surface reconstruction. Our findings establish a novel 2D Doniach phase diagram, distinct from conventional pressure-, doping-, or field-tuned heavy fermion systems. Similar phase diagrams, including heavy fermion superconductivity \cite{52,53}, are expected in other 2D moiré systems with coexisting flat and dispersive bands.

\begin{acknowledgments}
~ We thank Dr. Chao Zhang from the Instrumentation and Service Center for Physical Sciences (ISCPS) at Westlake University for technical support in data acquisition. This work was funded by National Natural Science Foundation of China (Grant No. 12550402, Grant No. 12274354, Grant No. 12574203, Grants No. 12474136 and No. 12174429), the National Key R\&D Program of China (Grant No. 2022YFA1402203), the Zhejiang Provincial Natural Science Foundation of China (Grant No. LR24A040003, XHD23A2001), Hangzhou Natural Science Foundation for Key Program (Grant No. 2025SZRJJ1093) and Westlake Education Foundation at Westlake University. K.W. and T.T. acknowledge support from the JSPS KAKENHI (Grant Numbers 21H05233 and 23H02052) and World Premier International Research Center Initiative (WPI), MEXT, Japan. Z.Z. acknowledges support from the European Union's Horizon 2020 research and innovation programme under the Marie-Sklodowska Curie grant agreement No 101034431.
\end{acknowledgments}

\section*{Data availability}
~ The data shown in the main figures and other findings that support this study are available from the corresponding authors upon reasonable request.

\balance

\onecolumngrid
\noindent\rule{\linewidth}{1.5pt}
\vspace{-1.2em}
\begin{center}
\textbf{\Large End Matter}
\end{center}
\vspace{0.2em}

\twocolumngrid

~ In this work, we also demonstrate that the observed isospin Pomeranchuk effect arises from the Kondo screening of localized moments by itinerant $c$-electrons, resulting in a heavy Fermi liquid ground state at low temperatures. Figure 6(a) and 6(b) show $R_{xx}$ transfer curves at temperatures ranging from 2 K to 150 K for two representative displacement field $D=0.1$ \si{V~nm^{-1}} and $D=0.9$ \si{V~nm^{-1}}, respectively. Additional temperature-dependent transfer curves for various displacement fields are provided in Supplemental Material Fig. S4 to Supplemental Material Fig. S6 \cite{SM}. At a small displacement field ($D=0.1$ \si{V~nm^{-1}}), resistance peaks emerge at $\nu=1,2,3$ upon cooling, attributable to the breaking of spin/valley degeneracy due to Coulomb interactions among localized moments \cite{54,55,56}. In contrast, at $\nu=3$ and larger displacement field $D=0.9$ \si{V~nm^{-1}}, we observe a completely opposite transport behavior: resistance peaks associated with correlated phase develop with increasing temperature and merge into a broad resistance peak around $\nu=0$ at temperatures above 54 K. This suggests that localized moments become detectable at higher temperature. The unusual behavior is identical to the recently confirmed isospin Pomeranchuk effect in MATBG \cite{57,58}, rhombohedral trilayer graphene superlattices \cite{59}, and trilayer-\ce{MoTe2}/\ce{WSe2} moiré superlattices \cite{60}. The Pomeranchuk effect, originally observed in the isotope \ce{^3He}, describes an entropy-driven liquid-to-solid transition upon heating, where the nuclear spin of atoms in the paramagnetic solid phase are disordered, leading to higher entropy compared to the liquid phase. In 2D moiré superlattices, the localized moments exhibit an analogous effect: a low-entropy Fermi liquid phase emerges at low temperatures, while a high-entropy, broadly isospin-polarized insulating phase dominates at high temperatures. In general, systems tend to maximize entropy at high temperatures by sampling all possible configurations. In heavy fermion materials, this is achieved by gaining free energy through the liberation of local moments, effectively rendering them free \cite{61}. 

\begin{figure}[t]
\centering
\includegraphics[width=0.52\columnwidth]{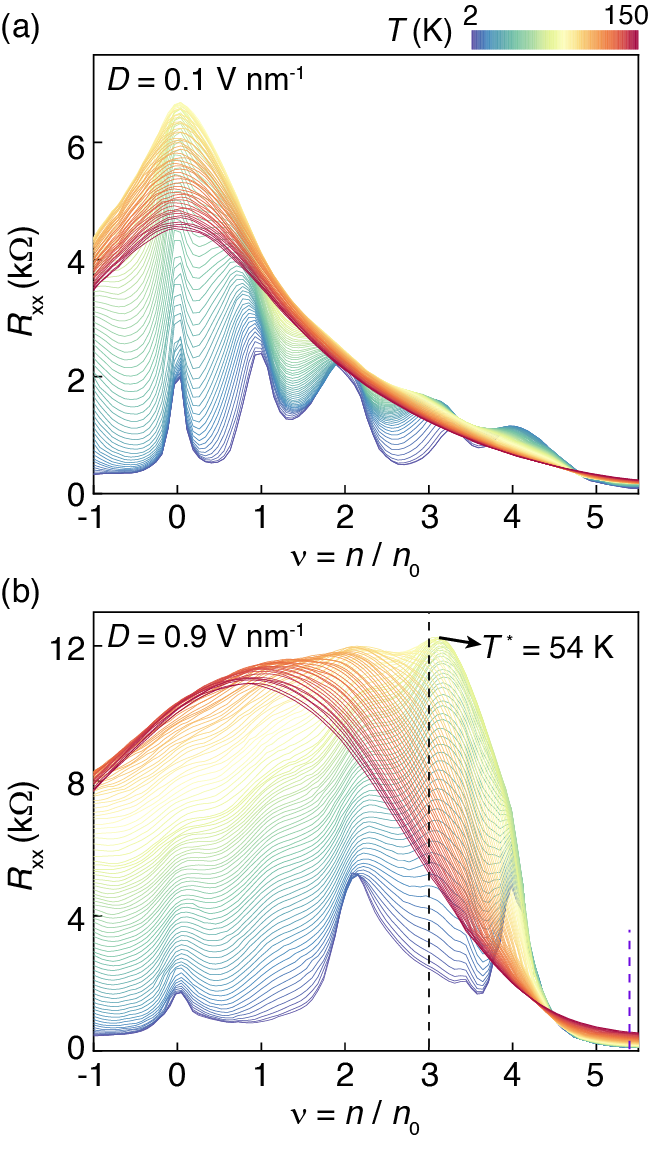}
\caption{\label{fig:fig6} Longitudinal resistance $R_{xx}$ versus $\nu$ measured at a small displacement field $D=0.1$ \si{V~nm^{-1}} (a) and a large displacement field $D=0.9$ \si{V~nm^{-1}} (b) across successive temperatures.}
\end{figure}

~ As shown in Fig. 7(a), we measured the ($\nu$, $D$) phase diagram of D2 at a base temperature of $T=$0.25 K. Compared to the measurement at $T=$2 K (see Supplemental Material Fig. S1 \cite{SM}), two prominent superconducting regions with zero resistance emerge at $\nu=-2-\delta$ and $\nu=2+\delta$ ($0<\delta<1$), which is consistent with previous reports. Hall measurements indicate that superconductivity is closely linked to flavor polarization at integer filling and to van Hove singularity at non-integer filling. For instance, at $D=0.36$ \si{V~nm^{-1}}, the $n_\text{H}$ increases linearly with filling factor until a charge carrier reset occurs at $\nu=1$. Following this flavor symmetry-breaking, $n_\text{H}$ exhibits divergent behavior at $\nu=1.8$, which corresponds to VHS. Superconductivity emerges after the correlated insulator gap at integer filling $\nu=2$ but vanishes at integer filling $\nu=3$ upon charge carrier resetting. Under higher displacement field, the superconductivity at $\nu=2+\delta$ is suppressed at the VHS. We also observe signature of superconductivity at $\nu=2-\delta$, which is similarly bounded by the VHS. One possible explanation for the incipient superconductivity is that pairing might occur at temperatures higher than those at which the zero-resistance state is detected. In conventional electron-phonon weak-coupling Bardeen-Cooper-Schrieffer theory, the density of state (DOS) peaks near a VHS, typically enhancing superconductivity. Together with prior work, our observations suggest an unconventional origin of the superconductivity in MATTG. In addition, we find a large in-plane upper critical magnetic field for the superconducting phase at optimal $\nu$ and $D$, which exceeds the Pauli limit. This finding may stimulate further investigations into the paring mechanism of graphene-based superconductors, such as spin-triplet paring.

% ========================================================

% ========================================================
\onecolumngrid
\par\vspace{0.4em}
\noindent\makebox[\textwidth][c]{\includegraphics[width=0.68\textwidth]{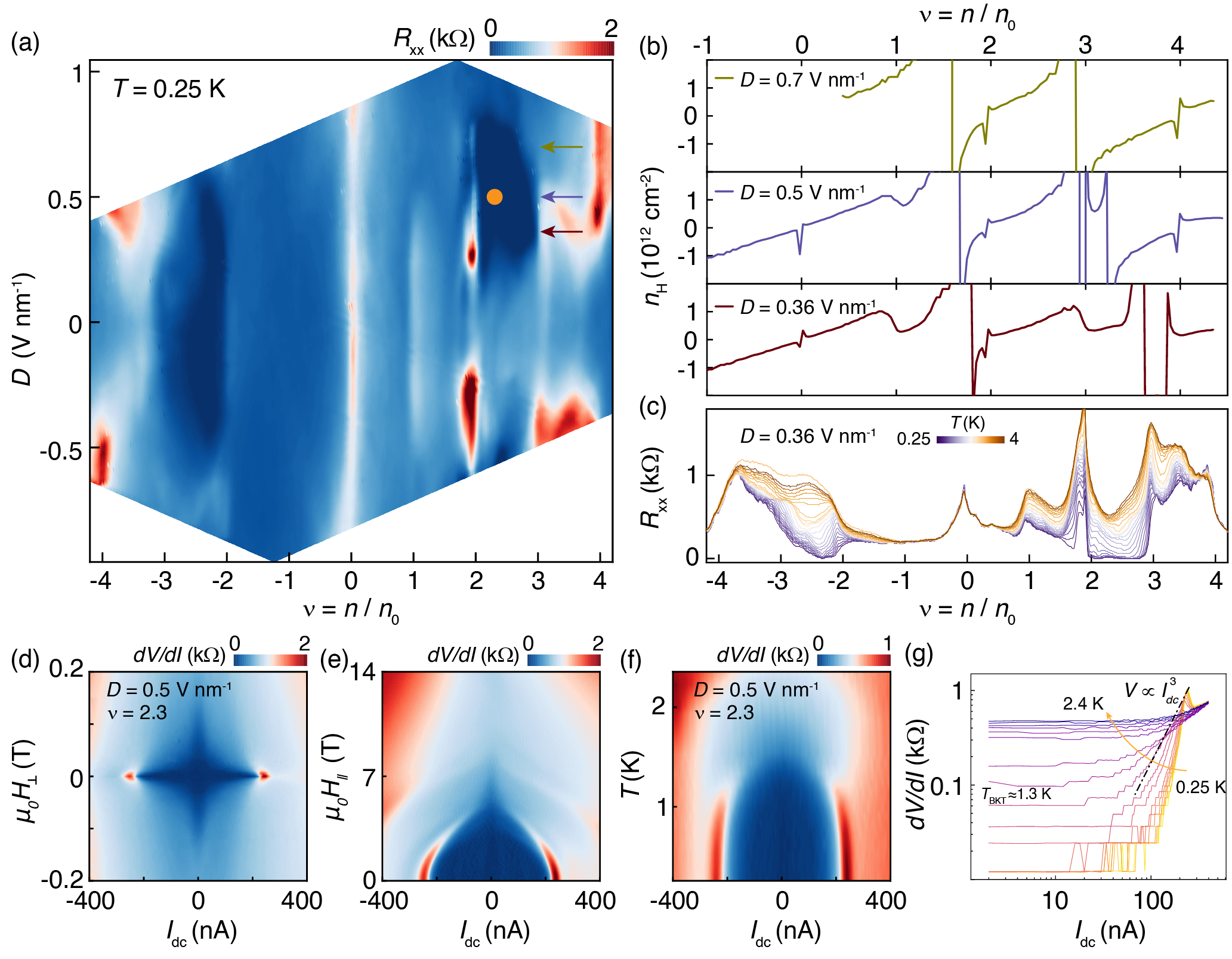}}
\par\vspace{3pt}
\refstepcounter{figure}\label{fig:fig7}
{\small\rmfamily FIG.~\thefigure.\ (a) $R_{xx}$ as a function of $\nu$ and $D$ at $T=$0.25 K. (b) Anti-symmetrized Hall carrier density $n_H$ as a function of $\nu$ at several $D$ values, corresponding to the horizontal arrows marked in (a). (c) $R_{xx}$ as a function of $\nu$ at $D=0.36$ \si{V~nm^{-1}}, measured at temperatures ranging from 0.25 K to 4 K. (d), (e) Critical current as a function of out-of-plane (d) and in-plane (e) magnetic field at $D=0.5$ \si{V~nm^{-1}} and $\nu=2.3$ [marked by the orange dot in (a)]. (f) Temperature-dependent differential resistance dV/dI at $D=0.5$ \si{V~nm^{-1}} and $\nu=2.3$. (g) $\mathrm{d}V/\mathrm{d}I$ versus d.c. current on a log-log scale. The dashed line corresponds to a $ V\propto I_{dc}^3$ power law dependence, indicating $T_\text{BKT}\approx1.3$ K.\par}

\end{document}